\documentclass[%
reprint,
showpacs,preprintnumbers,
 amsmath,amssymb,
 aip,
 graphicx,
apl,
]{revtex4-1}

\usepackage{graphicx}% Include figure files
\usepackage{dcolumn}% Align table columns on decimal point
\usepackage{bm}% bold math
\usepackage[mathlines]{lineno}% Enable numbering of text and display math
\begin{document}
%\preprint{Submitted to APL}

\title{Thermal conductivity of GeTe/Sb$_{2}$Te$_{3}$ superlattices measured by 
coherent phonon spectroscopy}
\author{Muneaki Hase}
\email{mhase@bk.tsukuba.ac.jp}
\affiliation{Institute of Applied Physics, University of Tsukuba, 1-1-1 Tennodai, 
Tsukuba 305-8573, Japan}
\author{Junji Tominaga}
\affiliation{Nanoelectronics Research institute, National Institute of Advanced Industrial Science and Technology, 
Tsukuba Central 4, 1-1-1 Higashi, Tsukuba 305-8562, Japan}

\date{\today}

\begin{abstract}
We report on evaluation of lattice thermal conductivity of GeTe/Sb$_{2}$Te$_{3}$ superlattice 
(SL) by using femtosecond coherent phonon spectroscopy at various lattice temperatures.  
The time-resolved transient reflectivity obtained in amorphous and crystalline GeTe/Sb$_{2}$Te$_{3}$ SL films 
exhibits the coherent $A_{1}$ optical modes at terahertz (THz) frequencies with picoseconds 
dephasing time. Based on the Debye theory, we calculate the lattice thermal conductivity, including 
scattering by grain boundary and point defect, umklapp process, and phonon resonant scattering.  
The results indicate that the thermal conductivity in amorphous SL is less temperature dependent, 
being attributed to dominant phonon-defect scattering. 
\end{abstract}


\maketitle

Phase change data storage technology offers high speed, rewritable, 
and reliable nonvolatile solid state memory, which may overcome current generation of 
Si-based memory technologies. In the phase change memory (PCM) materials, the switching 
between a high resistance amorphous and low resistance crystalline phases can be 
operated by optical means. 
One of the most common and reliable materials for the modern optical recording is 
Ge$_{2}$Sb$_{2}$Te$_{5}$ (GST), in which the phase transition between the crystalline and 
amorphous phases serves rewritable recording.\cite{Yamada91} Recently, extensive 
theoretical investigations on the mechanism of the phase change in GST have been made using 
molecular dynamics simulations.\cite{Akola07,Hegedus08} In addition, experimental studies 
using extended 
x-ray absorption fine structure (XAFS),\cite{Kolobov04} time-resolved x-ray absorption near-edge 
structure (XANES)\cite{Fons10} and Raman scattering measurements\cite{Andrikopoulos07} have 
examined local atomic arrangements in GST materials. 

One of the advantages of using GST films as the optical recording media is its ultra-high speed switching 
characteristics, whose time scale could be less than 1 nanosecond.\cite{Kolobov04} In the last decade, 
however, most of the literatures have studied nanosecond dynamics of the phase change in 
GST materials using nanosecond and picosecond laser (or electrical) pulses.\cite{Siegel04} 
Hence thermal properties of GST materials have been believed to govern the phase change in 
GST materials when it is promoted by laser heating. There, thermal conductivity ($\kappa$) is important to 
engineer the performance of the phase change,\cite{Lyeo06} such that lower thermal conductivity 
enables one to realize low power operation of the switching, where focused laser irradiation causes lattice 
heating.\cite{Lee09,Wong10} 

Coherent phonon spectroscopy (CPS) has recently been applied to 
GST materials of alloy \cite{Forst00} and superlatticed films,\cite{Hase09} and the related Sb$_{2}$Te$_{3}$ 
films.\cite{Li10} In their study, the 
observed local phonon modes in the amorphous GST films were found to be strongly damped 
modes, with its relaxation time of less than a few picoseconds due to the scattering by lattice 
defects.\cite{Forst00,Hase09}  The CPS on GST compounds, however, have not yet 
been applied to investigate thermal properties, although Wang {\it et al.} have recently 
proposed to use CPS as a powerful method to estimate lattice thermal conductivity.\cite{Wang09} 
 
In this paper, we present detailed analysis on the ultrafast dynamics of coherent optical 
phonons in GeTe/Sb$_{2}$Te$_{3}$ SLs at low and room temperatures to investigate 
lattice thermal conductivity. Based on the Debye theory, we calculate the lattice thermal 
conductivity, including various phonon scattering processes, where the relaxation rate and the frequency 
of the observed coherent local modes are included in the model.  The results indicate in amorphous ({\it a}--) 
GeTe/Sb$_{2}$Te$_{3}$ SL that $\kappa$ $\approx$ 0.3--0.4 Wm$^{-1}$K$^{-1}$ at T $\geq$ 100 K, 
while in the crystalline ({\it c}--) GeTe/Sb$_{2}$Te$_{3}$ SL 
$\kappa$ is strongly temperature dependent and $\kappa$ $\approx$ 2.0 Wm$^{-1}$K$^{-1}$ at 300 K. 

We have chosen GeTe/Sb$_{2}$Te$_{3}$ SL as a sample after the proposal of a class of superlattice-like 
GeTe/Sb$_{2}$Te$_{3}$.\cite{Chong06} Significantly lower SET and RESET programming 
current for the SL cells has already been discovered \cite{Chong06} and thus 
GeTe/Sb$_{2}$Te$_{3}$ SL will be a potential candidate for the future PCM devises. 
The samples used in the present study were thin films of GeTe/Sb$_{2}$Te$_{3}$ SL fabricated 
using a helicon-wave RF magnetron sputtering machine on Si (100) substrate. The thickness of the films was 20 nm. 
The annealing of the as-grown SL films at 503 K (230 $^{\circ}$C) for ten minutes changed 
the amorphous into the crystalline states.\cite{Tominaga08} The TEM measurements 
confirmed that the {\it c}--GeTe/Sb$_{2}$Te$_{3}$ SL films have layered structures with clear interfaces. 

A reflection-type pump-probe measurements using a mode-locked Ti:sapphire laser 
(pulse width = 20 fs and a central wavelength = 850 nm) was employed at the temperature range 
of 5 - 300 K. 
The average power of the pump and probe beams were fixed at 120 and 3 mW, 
respectively, from which we estimated the pump fluence to be 284 $\mu$J/cm$^{2}$ at 120 mW. 
The excitation of the GST-SL films with the 850 nm (= 1.46 eV) laser pulse generates 
nonequilibrium carriers across the narrow band gap of $\approx$ 0.5 - 0.7 eV.\cite{Lee05} 
The transient reflectivity (TR) change ($\Delta R/R$) was measured as a function of the time 
delay. 

\begin{figure}
\includegraphics[width=7.5cm]{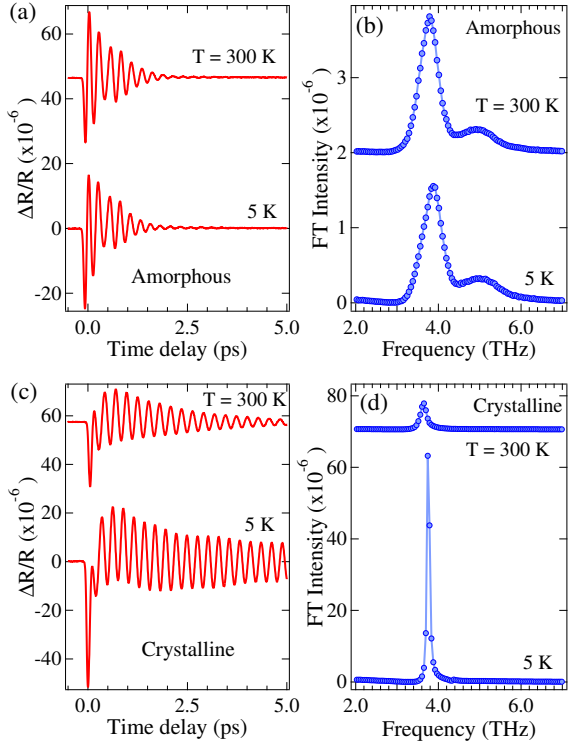}
\caption{(Color online). (a) (c) The time-resolved TR signal observed at 5 and 300 K in the 
{\it a--} and {\it c--}GeTe/Sb$_{2}$Te$_{3}$ SL films, respectively. (b) (d) The FT spectra 
obtained from the time-domain data in (a) and (c).  
}
\label{Fig1}
\end{figure}
Figures 1(a) and (c) show the time-resolved TR signal ($\Delta R/R$) observed at 5 and 300 K in 
GeTe/Sb$_{2}$Te$_{3}$ SL films with the amorphous and crystalline phases, respectively. After the transient 
electronic response due to the excitation of nonequilibrium carriers at the time delay zero, 
coherent oscillations with several picoseconds relaxation time appear. Fourier transformed (FT) 
spectra in Figs. 1(b) and (d) are obtained from the time-domain data, in which two broad 
peaks are observed at $\approx$ 5.1 THz and $\approx$ 3.78 THz in amorphous film, while 
a sharp peak at 3.68 THz is observed in crystalline film at 300 K. 
These peaks in the amorphous film were assigned to the A$_{1}$ optical modes due to 
the local GeTe$_{4}$ unit (3.78 THz peak),\cite{Forst00,Hase09} and that due to the local 
pyramidal SbTe$_{3}$ unit (5.1 THz peak) \cite{Andrikopoulos07,Hase09}. The red shift of the local 
$A_{1}$ mode frequency in the crystalline phase (3.78 THz $\Rightarrow$ 3.68 THz) 
has been attributed to the local structural change from tetrahedral GeTe$_{4}$ into octahedral 
GeTe$_{6}$ species.\cite{Hase09} 
It is to be noted that the zone-folding modes of the acoustic dispersion \cite{Bartels99} cannot be detected in our 
measurements because the SL period ($d$ $\approx$ 5{\AA}) of GeTe/Sb$_{2}$Te$_{3}$ film 
is an order of the lattice constant. 

To investigate the effect of these local phonon modes on lattice thermal conductivity, the 
parameters of the coherent local phonons (the frequency and the relaxation rate) are used to 
compute the lattice thermal conductivity based on the Debye theory, combined with the 
resonant scattering model.\cite{Pohl62,Wang09} Lattice thermal conductivity is expressed 
as,\cite{Callaway} 
\begin{eqnarray}
\kappa (T) 
& =  & \frac{1}{3}C_{V}v^{2}\tau_{c} \\ 
& =  & \frac{k_{B}}{2 \pi^{2} v}\Bigl(\frac{k_{B}T}{\hbar} \Bigr)^{3}
{\displaystyle\int\nolimits_{0}^{\Theta_{D}/T}}
\frac{x^{4}e^{x}}{\tau_{c}^{-1}(e^{x}-1)^{2}}dx\text{ \ ,}
\end{eqnarray}
where $x = \hbar \omega/k_{B}T$, $C_{V}$ is the lattice specific heat, $v$ the sound velocity, 
$\Theta_{D}$ the Debye temperature,\cite{Kuwahara07} $k_{B}$ the Boltzmann constant, $\omega$ the phonon 
frequency, $\tau_{c}$ the acoustic phonon relaxation time, whose inverse (relaxation rate) can 
be given by contributions from various scattering mechanisms:\cite{Pohl62,Wang09} 
\begin{equation}
\tau_{c}^{-1} =\frac{v}{L} + A\omega^{4} + B\omega^{2}T e^{- \Theta_{D}/3T} + 
\frac{C \omega^{2}}{(\Omega^{2}-\omega^{2})^{2}}\text{ \ ,}
\end{equation}
where $L$, $A$, $B$, and $C$ characterize grain boundary, phonon-defect scattering, 
phonon-phonon umklapp scattering, and phonon resonant scattering, respectively. 
$\Omega$ is the optical phonon frequency observed in the CPS and the last term in Eq. (3) 
represents the resonant scattering between the localized optical modes and acoustic phonon modes. 

 \begin{table*}
  \caption{Parameters used in Eqs. (2) and (3).  For the {\it a}--GeTe/Sb$_{2}$Te$_{3}$ SL $C_{1}$ and $C_{1}$ represent the resonant scattering 
  coefficient due to the A$_{1}$ local modes at 3.78 THz and 5.1 THz, respectively, while for the {\it c}--GeTe/Sb$_{2}$Te$_{3}$ SL $C_{1}$ represents that at 3.68 THz. $^{a}$from 
  Ref. \cite{Kuwahara07} and $^{b}$from Ref. \cite{Lyeo06}.}
\begin{ruledtabular}
  \begin{tabular}{cccccccc}
     Samples & $\Theta_{D}$(K) & $v$(m/s) & $L$(nm) & $A$(10$^{-43}$ $s^{3}$) & $B$(10$^{-18}$ $sK^{-1}$) & $C_{1}$(10$^{38}$ $s^{-3}$) & $C_{2}$(10$^{38}$ $s^{-3}$)\\
    \hline
  {\it a}--GeTe/Sb$_{2}$Te$_{3}$ SL & 250$^{a}$ & 2250$^{b}$ & 10.0 & 40.0 & 40.0 & 20.0 & 20.0 \\
  {\it c}--GeTe/Sb$_{2}$Te$_{3}$ SL & 300$^{a}$ & 3190$^{b}$ & 100.0 & 4.0 & 4.0 & 2.0 & - \\
  \end{tabular}
  \end{ruledtabular}
\end{table*}
From the low temperature limit of the relaxation rate of the coherent A$_{1}$ modes (0.253 ps$^{-1}$ 
for the amorphous and 0.026 ps$^{-1}$ for the crystalline phase),\cite{Hase09} we can estimate 
the ratio of the phonon-defect scattering rate in the amorphous to the crystalline $A_{a}/A_{c}$ to be $\approx$ 10 for the 
GeTe/Sb$_{2}$Te$_{3}$ SL film. The same ratio of $B_{a}/B_{c}$ = $C_{a}/C_{c}$ = 10 has been applied in 
the simulation. We take the resonant phonon frequency ($\Omega$) 
at 300 K from the FT spectra in Fig. 1. It is to be noted that the choice of the optical phonon frequency at 
different temperatures does not significantly affect the results of the thermal conductivity, but the 
coefficient of the phonon resonant scattering ($C$) is more sensitive to the value of $\kappa$. 
The magnitudes of all the parameters ($L$, $A$, $B$, and $C$) are determined as listed in Table 
I to give the experimental value of $\kappa$ for the {\it a}--GeTe/Sb$_{2}$Te$_{3}$ SL ($\kappa$ $\approx$ 
0.33 Wm$^{-1}$K$^{-1}$ at 300 K).\cite{Simpson} 

\begin{figure}
\includegraphics[width=7.5cm]{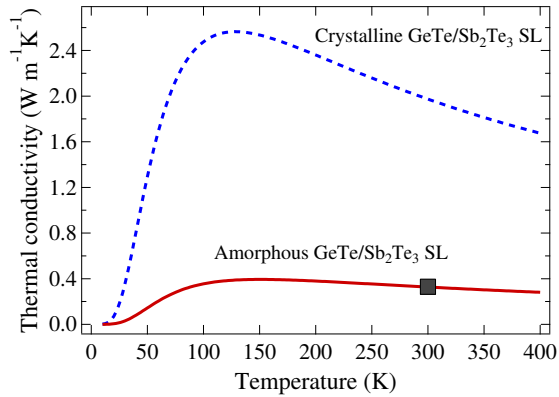} 
\caption{(Color online). Lattice thermal conductivity of the SL films as a function of lattice temperature 
calculated by Eqs. (2) and (3). The closed square at 300 K represents the experimental datum 
for the {\it a}--GeTe/Sb$_{2}$Te$_{3}$ SL obtained by thermal reflectance measurements.\cite{Simpson} 
}
\label{Fig2}
\end{figure}
As shown in Fig. 2, comparing the thermal conductivity obtained for the SL films in different phases, 
we found that  the thermal conductivity in {\it a}--GeTe/Sb$_{2}$Te$_{3}$ SL is less temperature dependent, 
being due to dominant contribution from the phonon-defect scattering.\cite{Hase09,Hase00} On the contrary, 
in the crystalline phase thermal conductivity is strongly temperature dependent, being attributed to 
significant contribution from umklapp and phonon resonant scatterings, both of which are related to 
the phonon dispersion curves and therefore they are significantly temperature dependent.\cite{Klemens} 
We note further that the thermal conductivity obtained in the SL films is high compared to the 
conventional GST alloy films; $\kappa$ $\approx$ 0.2 Wm$^{-1}$K$^{-1}$ for the amorphous 
and $\kappa$ $\approx$ 0.4 Wm$^{-1}$K$^{-1}$ for the crystalline (cubic) phases.\cite{Lyeo06} 
The higher thermal conductivity while the lower operation current found in GeTe/Sb$_{2}$Te$_{3}$ SL 
films, suggests that the phase change in the SL films under the irradiation of ultrashort laser pulses would not be promoted by thermal process, 
but rather by nonthermal process, which has recently been observed in sub-picosecond time scale.\cite{Makino11} 

In conclusion, our results on ultrafast coherent phonon spectroscopy have illustrated temperature 
dependence of lattice thermal conductivity in GeTe/Sb$_{2}$Te$_{3}$ SL films. These data show that the Debye model, 
including scatterings by grain boundary and point defect, umklapp process, and phonon resonant 
scattering, well reproduces the experimental value of thermal conductivity measured by using 
thermo-reflectance.  The thermal conductivity in the {\it a}--SL film is less temperature dependent, 
due to the dominant phonon-defect scattering, while in the {\it c}--SL 
it is strongly temperature dependent because of the main contributions from umklapp and phonon resonant 
scatterings. We argue the higher thermal conductivity in the SL films implies that the phase change 
in GeTe/Sb$_{2}$Te$_{3}$ SL under the irradiation of ultrashort laser pulses is not promoted by {\it thermal} process, i.e., lattice heating, but rather by {\it nonthermal} 
process, i.e., coherent lattice excitation, because the thermal process requires lower thermal conductivity.\cite{Lee09,Wong10}  

The authors thank Y. Miyamoto for the assistance at the early stage of the experiments. 
This work was supported in part by PRESTO-JST, KAKENHI-22340076 from MEXT, Japan and "Innovation Research 
Project on Nanoelectronics Materials and Structures -- Research and development of superlatticed 
chalcogenide phase--change memory based on new functional structures" from METI, Japan. 

%\vspace{1.5 cm}

%\pagebreak

\end{document}